\documentclass[12pt]{article}

\newcommand{\be}{\begin{equation}}
\newcommand{\ee}{\end{equation}}
\newcommand{\bi}[1]{\vspace{-3mm} \bibitem{#1}}

\usepackage{epsfig,amsmath,amssymb,graphics,graphicx} 

\begin{document}
\begin{center}

{\it International Journal of Modern Physics B. Vol.27. No.9. (2013) 1330005.}

\vskip 7mm
{\bf \large Review of Some Promising Fractional Physical Models} \\

\vskip 3mm
{\bf \large Vasily E. Tarasov} \\
\vskip 3mm

{\it Skobeltsyn Institute of Nuclear Physics,\\ 
Lomonosov Moscow State University, Moscow 119991, Russia} \\
{E-mail: tarasov@theory.sinp.msu.ru} \\

\begin{abstract}
Fractional dynamics is a field of study in physics and 
mechanics investigating the behavior of objects and systems 
that are characterized by power-law non-locality, 
power-law long-term memory or fractal properties
by using integrations and differentiation of non-integer orders, 
i.e., by methods of the fractional calculus.
This paper is a review of physical models that look 
very promising for future development of fractional dynamics.
We suggest a short introduction to fractional calculus as a theory
of integration and differentiation of non-integer order.
Some applications of integro-differentiations 
of fractional orders in physics are discussed. 
Models of discrete systems with memory,  
lattice with long-range inter-particle interaction, 
dynamics of fractal media are presented. 
Quantum analogs of fractional derivatives and model of
open nano-system systems with memory are also discussed.
\end{abstract}

\end{center}



\newpage
\tableofcontents

\newpage
\section{Introduction}

In this paper we review some applications of
fractional calculus and fractional differential equations in physics and mechanics.
The interest in such applications has been growing continually during the last years. 
Fractional calculus is a theory of integrals and 
derivatives of any arbitrary real (or complex) order. 
It has a long history \cite{Ross,MKM,OS} from 30 September 1695, 
when the derivatives of order $\alpha=1/2$ has 
been described by Leibniz in a letter to L'Hospital. 
Therefore this date can be regarded as the birthday of Fractional Calculus. 
We can probably think that Joseph Liouville was a first 
in application of fractional calculus in physics \cite{Lutzen}. 
The fractional differentiation and fractional integration 
go back to many great mathematicians such as 
Leibniz, Liouville, Riemann, Abel, Riesz, Weyl. 


All of us are familiar with derivatives and integrals, like first order
\[ f^{\prime}(x)=D^1_x f(x) = \frac{d}{dx} f(x) , 
\quad (I^{1} f)(x) = \int^x_0 d x_{1} \, f(x_1) . \]
and the $n$-th order  
\[ f^{(n)}(x)=D^n_x f(x) = \frac{d^n}{dx^n} f(x) , \]
\[ (I^{n} f)(x) = \int^x_0 d  x_{1} \int^{x_1}_0 dx_{2} ... \int^{x_{n-1}}_0 dx_n \, f(x_n) , \]
where $n$ is integer number $n=1,2,...$, i.e. $n \in \mathbb{N}$.

Mathematicians consider 
the non-integer order of the integrals and derivatives from 1695, 
\[ f^{(\alpha)}(x)=D^{\alpha}_x f(x) =  \  ? , \quad (I^{\alpha} f)(x) = ? \ , \]
where $\alpha \in \mathbb{R}$ or $\alpha \in \mathbb{C}$. 

At this moments there are international journals such as
"Fractional Calculus and Applied Analysis", "Fractional Differential Calculus", "Communications in Fractional Calculus",
which are dedicated entirely to the fractional calculus.


The first book dedicated specifically to fractional calculus
is the book by Oldham and Spanier \cite{OS} published in 1974.
There are two remarkably comprehensive encyclopedic-type monographs.
The first such monographs is written by Samko, Kilbus and Marichev \cite{SKM} and 
it was published in Russian in 1987 and in English in 1993.
In 2006 Kilbas, Srivastava and Trujillo published a very important and 
other remarkable book \cite{KST}, where one can find a modern encyclopedic, 
detailed and rigorous theory of fractional differential equations.
It should be noted the books on fractional differential equation by Podlubny \cite{Podlubny} and 
an introduction to fractional calculus for physicists by Herrmann \cite{Herrmann}.
There exist mathematical monographs devoted 
to special questions of fractional calculus. 
For example, these include the book by McBride \cite{Bride1} 
published in 1979 (see also \cite{Bride2}),
the work by Kiryakova \cite{Kiryakova} of 1993. 
The fractional integrals and potentials are described in the monograph by Rubin \cite{Rubin},
the univalent functions, fractional calculus and their applications are described in
the volume edited by Srivastava and Owa \cite{SOwa}.
Fractional differentiation inequalities are described 
in the book by Anastassiou \cite{Anastassiou} published in 2009.


The physical applications of fractional calculus 
to describe complex media and processes 
are considered in the very interesting volume edited by 
Carpintery and Mainardi \cite{CM} publish  in 1997. 
Different physical systems are described in
the papers of volumes edited by Hilfer \cite{Hilfer} in 2000,
and the edited volume of Sabatier, Agrawal and Tenreiro Machado 
\cite{SATM} published in 2007. 
The most recent volumes on the subject of application of
fractional calculus is the volumes edited by 
Luo and Afraimovich \cite{LA} in 2010, and 
by Klafter, Lim, and Metzler \cite{KLM} in 2011.
The book by West, Bologna, and Grigolini \cite{WBG}
published in 2003 is devoted to physical application 
of fractional calculus to fractal processes. 
The first book devoted to application of fractional calculus
to chaos is the book by Zaslavsky \cite{Zaslavsky2} published in 2005.
The interesting book by Mainardi \cite{Mainardi} devoted to 
application of fractional calculus in dynamics of viscoelastic materials.
The books dedicated specifically to application of derivatives and integrals
with non-integer orders in theoretical physics 
are the remarkable books by Uchaikin \cite{Uchaikin,Uchaikin3}, 
and the monographs by Tarasov \cite{TarasovSpringer,TarasovRCD}.
We also note a new book by Uchaikin and Sibatov \cite{Uchaikin2}
devoted to fractional kinetics in solids. 


Due to the fact that there are many books and reviews on 
application of fractional calculus to describe physical processes and systems,
it is almost impossible in this review to cover all areas of current research 
in the field of fractional dynamics. 
Therefore we must to choose some of the areas in this field. 
We have chosen areas of fractional dynamics that can be considered as 
the most perspective directions of research in my opinion.
These areas are not related to a simple extension of equations 
with derivatives of integer order to non-integer.
We consider fractional models that give relationships between different types of equations describing
apparently the system and processes of various types.
In addition, we think that these areas and models
can give new prospects for a huge number of fundamentally new results 
in the construction of mathematical methods for the solution of physical problems, 
and in the description of new types of physical processes and systems.

As a first type of models, we consider discrete maps with memory that are equivalent to the
fractional differential equations of kicked motions.
These models are promising since an approximation for fractional derivatives 
of these equations of motion is not used.
This fact allows us to study the fractional dynamics by computer simulations
without approximations. It allows us to find and investigate 
a new type of chaotic motion and a new type of attractors.

As a second type of promising fractional models, we consider the discrete systems (or media) 
with long-range interaction of particles,
and continuous limits of these systems such that
equations of motion with long-range interaction
are mapped into continuous medium equations with the fractional derivatives. 
As a result we have microscopic model for fractional dynamics of complex media.

The third type of models is related to the description of the fractional dynamics 
by microscopic models of open quantum systems which interacts with its environment.
We give an example that demonstrate that time fractional dynamics and
a fractional differential equation of motion can be connected
with the interaction between the system and its environment with power-law spectral density. 

We also consider fractional models that allow us to describe specific properties of fractal media dynamics;
quantum analogs of fractional derivative with respect to coordinate and momentum;
the importance of self-consistent formulation of fractal vector calculus 
and exterior calculus of differential forms
that are not yet fully implemented.
This review starts with a short introduction to the fractional calculus.

 
\section{Derivatives and integrals of non-integer orders}

There are many different definitions of fractional integrals and
derivatives of non-integer orders. 
The most popular definitions are based on the following.

1) A generalization of Cauchy's differentiation formula; 

2) A generalization of finite difference; 

3) An application of the Fourier transform. 

We should note that many usual properties of the ordinary derivative $D^n$ 
are not realized for fractional derivative operators $D^{\alpha}$. 
For example, a product rule, chain rule, semi-group property 
have strongly complicated analogs for the operators $D^{\alpha}$.


\subsection{A generalization of Cauchy's differentiation formula}

Let $G$ be an open subset of the complex plane $\mathbb{C}$, 
and $f : G \to \mathbb{C}$ is a holomorphic function.
Then we have the Cauchy's differentiation formula
\be \label{Cauchy}
 f^{(n)}(x) = {n! \over 2\pi i} \oint_L {f(z) \over (z-x)^{n+1}} \, dz. 
\ee
A generalization of (\ref{Cauchy}) has been suggested by
Sonin (1872) and Letnikov (1872) in the form
\be \label{SL}
D^{\alpha}_x f(x) = {\Gamma(\alpha+1) \over 2\pi i} 
\oint_L  {f(z) \over (z-x)^{\alpha+1}} \, dz, 
\ee
where $\alpha \in \mathbb{R}$ and $\alpha \not= -1,-2,-3,..$.
See Theorem 22.1 in the book by Samko, Kilbas, and Marichev \cite{SKM}.
Expression (\ref{SL}) is also called Nishimoto derivative \cite{Nishimoto}. 
More correctly it should be called Sonin-Letnikov derivative.

\subsection{A generalization of finite difference}

It is well-known that derivatives of integer orders $n$ can be defined by the finite differences.
The differentiation of integer order $n$ can be defined by
\[ D^n_x f(x) = \lim_{h \to  0} \frac{\Delta^n_h f(x)}{h^n} , \]
where $\Delta^n_h$ is a finite difference of integer order $n$ that is defined by
\be \label{dif-2}
\Delta^n_h f(x) = \sum_{k = 0}^{n} (-1)^k \binom{n}{k} f(x - k h) .
\ee
The difference of non-integer order $\alpha>0$ is defined by the infinite series
\be \label{dif-3}
\Delta^{\alpha}_h f(x) = 
\sum_{k = 0}^{\infty} (-1)^k \binom{\alpha}{k} f(x - k h) ,
\ee
where the binomial coefficients are
\[ \binom{\alpha}{\beta} = 
\frac{(\Gamma(\alpha+1)}{\Gamma(\beta+1) \Gamma(\alpha - \beta +1)} . \]

The left- and right-sided Gr\"unwald-Letnikov (1867,1868) derivatives 
of order $\alpha>0$ are defined by
\be \label{GL-ID}
^{GL}D^{\alpha}_{x \pm} f(x) = 
\lim_{h \to  0} \frac{\Delta^{\alpha}_{\mp h} f(x)}{h^{\alpha}} . \ee

It is interesting that series (\ref{dif-3}) 
can be used for $\alpha<0$ 
and equation (\ref{GL-ID}) defines Gr\"unwald-Letnikov fractional integral if 
\[ |f(x)|  < c (1+|x|)^{-\mu} , \quad \mu> |\alpha| . \]
Then (\ref{GL-ID}) can be represented by 
\[ ^{GL}D^{\alpha}_{x \pm} f(x) =  
\frac{\alpha}{ \Gamma(1-\alpha)} 
\int^{\infty}_0  \frac{f(x)-f(x \mp z)}{z^{\alpha+1}} dz  \]
if $f(x) \in L_p(\mathbb{R})$, 
where $1<p<1/\alpha$ and $0<\alpha<1$.

\subsection{A generalization by Fourier transform}

If we define the Fourier transform operator ${\cal F}$ by
\be \label{Fourier-operator} ({\cal F} f)(\omega)=
\frac{1}{2\pi} \int^{+\infty}_{-\infty} f(t) e^{-i\omega t} d t , \ee 
then the Fourier transform of derivative of integer order $n$ is
\[ ({\cal F} D^n_x f)(\omega)=(i\omega)^n ({\cal F}f)(\omega) . \]
Therefore we can define the derivative of integer order $n$ by
\[ D^n_x f(x) = {\cal F}^{-1} \{ (i\omega)^n ({\cal F} f)(\omega) \} . \]

For $f(t) \in L_1(\mathbb{R})$, the left- and right-sided
Liouville fractional derivatives and integrals can be defined 
(see Theorem 7.1 in \cite{SKM} and  Theorem 2.15 in \cite{KST})
by the relations 
\be \label{LD-def}
(D^{\alpha}_{\pm}f)(x)=
{\cal F}^{-1} \Bigl( (\pm i\omega)^{\alpha} ({\cal F}f)(\omega) \Bigr), \ee
\be \label{LI-def}
(I^{\alpha}_{\pm} f)(x)= {\cal F}^{-1} 
\Bigl( \frac{1}{ (\pm i\omega)^{\alpha} } ( {\cal F} f )(\omega) \Bigr) , \ee
where $0<\alpha<1$ and
\[ (\pm i\omega)^{\alpha} = |\omega|^{\alpha} 
\exp \Bigl( \pm sgn(\omega) \frac{i \, \alpha \, \pi}{2} \Bigr) . \]


The Liouville fractional integrals (\ref{LI-def}) can be represented by
\be 
(I^{\alpha}_{ \pm }f)(x) =\frac{1}{\Gamma(\alpha)} 
\int^{\infty}_{0} z^{\alpha-1} f(x \mp z ) dz . 
\ee
The Liouville fractional derivatives (\ref{LD-def}) are
\be (D^{\alpha}_{ \pm }f)(x) = D^n_x (I^{n-\alpha}_{ \pm }f)(x)  = \frac{1}{\Gamma(n-\alpha)} 
\frac{d^n}{dx^n} \int^{\infty}_{0} z^{n-\alpha-1} f(x \mp z ) dz , \ee
where $n=[\alpha]+1$. 


We can define the derivative of fractional order $\alpha$ by 
\be ^CD^{\alpha}_{\pm} f(t) = I^{n-\alpha}_{ \pm } (D^n_t f)(t)
= \frac{1}{\Gamma(n-\alpha)} 
\int^{\infty}_{0} z^{n-\alpha-1} D^n_x f(x \mp z ) dz , \ee
where $n=[\alpha]+1$. 
It is the Caputo derivative of order $\alpha$ \cite{KST}.
For $x \in [a,b]$ the left-sided Caputo fractional derivative 
of order $\alpha >0$ is defined by
\be \label{4-1}
\,  _a^CD^{\alpha}_t f(t)= \, _aI^{n-\alpha}_t D^n_t f(t) =
\frac{1}{\Gamma(n-\alpha)} \int^t_a 
\frac{ d\tau \, D^n_{\tau}f(\tau)}{(t-\tau)^{\alpha-n+1}} ,
\ee
where $n-1 < \alpha <n$, and $_aI^{\alpha}_t$ is 
the left-sided Riemann-Liouville fractional integral 
of order $\alpha >0$ that is defined by
\[ _aI^{\alpha}_t f(t)=\frac{1}{\Gamma(\alpha)} 
\int^t_a \frac{f(\tau) d \tau}{(t-\tau)^{1-\alpha}} , \quad (a<t). \]


Note that the Riemann-Liouville fractional derivative has some notable 
disadvantages in applications such as 
nonzero of the fractional derivative of constants, 
\[ _0D^{\alpha}_t C=\frac{t^{-\alpha}}{\Gamma(1-\alpha)} C , \]
which means that dissipation 
does not vanish for a system in equilibrium.  
The Caputo fractional differentiation of a constant results in zero
\[ _0^CD^{\alpha}_t C=0 . \]

The desire to use the usual initial value problems 
\[ f(t_0)=C_0, \quad (D^1_tf)(t_0)=C_1, \quad (D^2_tf)(t_0)=C_2, ... \]
lead to the application Caputo fractional derivatives 
instead of the Riemann-Liouville derivative.

The Riemann-Liouville and Caputo derivative are connected \cite{KST}.
Let $f(t)$ be a function for which 
the Caputo derivatives of order $\alpha$ exist
together with the Riemann-Liouville derivatives.
Then these fractional derivatives are connected by the relation
\be \label{4-CRLa}
_a^CD^{\alpha}_t f(t)= \, _aD^{\alpha}_t f(t)-
\sum^{m-1}_{k=0}
\frac{(t-a)^{k-\alpha}}{\Gamma(k-\alpha+1)} f^{(k)}(a) . \ee 
The second term of the right hand side of equation (\ref{4-CRLa}) 
regularizes the Caputo fractional derivative to avoid 
the potentially divergence from singular integration at $t=0$. 

\subsection{Some unusual properties of fractional derivatives}

Let us demonstrate the unusual properties of derivatives of non-integer orders
by using the Riemann-Liouville derivatives.

1) Semi-group property does not hold
\be \label{UUP-1}
(D^{\alpha}_{a+}  D^{\beta}_{a+} f)(x) = (D^{\alpha+\beta}_{a+} f )(x) -
\sum^{[\beta]+1}_{k=1} (D^{\beta-k}_{a+}  f)(a) 
\frac{(x-a)^{-\alpha-k}}{\Gamma(1-\alpha-k)} \ee
for $f(x) \in L_1(a,b)$ and $(I^{n-\alpha}_{a+} f)(x) \in AC^{n}[a,b]$,
(see equation 2.1.42 in \cite{KST}).

As a consequence, in general we have
\[ D^{\alpha}_{a+}  D^{\alpha}_{a+}  \not= D^{2\alpha}_{a+}  . \]

2) The derivative of the non-zero constant is not equal to zero
\be \label{UUP-2}
(D^{\alpha}_{a+} 1)(x) =\frac{(x-a)^{-\alpha}}{\Gamma(1-\alpha)}  . \ee

3) The initial conditions for differential equation with Riemann-Liouville derivative 
differ from the conditions for ordinary differential equations of the integer order 
\be \label{UUP-3}
( \, _0D^{\alpha - k}_t x)(0+) =c_k, \quad k=1,...,n . \ee
For example, conditions (\ref{UUP-3}) for $1<\alpha<2$ give
\[ ( \, _0D^{\alpha-1}_t x)(0+)=c_1, \quad 
( \, _0D^{\alpha-2}_t x)(0+)=( \, _0I^{2-\alpha}_t x)(0+) = c_2 . \]

4) Representation in the form of an infinite series of derivatives of integer orders
\be \label{UUP-4}
(D^{\alpha}_{a+} f)(x) = \sum^{\infty}_{n=0}  
\frac{\Gamma(\alpha+1)}{\Gamma(k+1) \Gamma(\alpha-k+1)}   
\frac{(x-a)^{k-\alpha}}{\Gamma(k-\alpha+1)}  D^n_x f(x)  \ee
for analytic (expandable in a power series on the interval) functions on $(a,b)$,
(see Lemma 15.3 in \cite{SKM}).

5) A generalization of the classical Leibniz rule 
\[ D^n(fg) = \sum^n_{k=0} \binom{n}{k} f^{(n-k)} g^{(k)}  \]
from integer $n$ to fractional $\alpha$ contains an infinite series
\be \label{UUP-5}
(D^{\alpha}_{a+} (f g)) (x) =
\sum^{\infty}_{k=0}  \frac{\Gamma(\alpha+1)}{\Gamma(k+1) \Gamma(\alpha-k+1)} 
 (D^{\alpha-k}_x f)(x) \, D^{k}_x g  \ee
for analytic functions on $(a,b)$ (see  Theorem 15.1 in \cite{SKM}).
The sum is infinite and contains integrals of fractional order (for $k>[\alpha]+1$).

6) The increasing complexity of the Newton-Leibniz equation for
\[ ( I^{\alpha}_{a+} D^{\alpha}_{a+} f)(x) =
f(x) - \sum^{n}_{k=1} \frac{(x-a)^{\alpha-k}}{\Gamma(\alpha-k+1)} 
\, (D^{n-k}_x I^{n-\alpha}_{a+}f)(a)  \]
for $f(x) \in L_1(a,b)$, $(I^{n-\alpha}f)(x) \in AC^n[a,b]$,  
(see equation 2.1.39 in \cite{KST}).

For $0<\alpha \le 1$, we have
\[ ( I^{\alpha}_{a+} D^{\alpha}_{a+} f)(b) =
f(b) - \frac{(b-a)^{\alpha-1}}{\Gamma(\alpha)} 
(I^{1-\alpha}_{a+} f)(a)  . \]
For $n \in \mathbb{N}$ 
\[ ( I^n_{a+} D^n_{a+} f)(x) =
f(b) - \sum^{n-1}_{k=0} \frac{f^{(k)}(a)}{k!} (b-a)^k .\]
\[ ( I^1_{a+} D^1_{a+} f)(b) = f(b) - f(a) , \]
(see equation 2.1.41 in \cite{KST}).

7) For the fractional derivatives there is an analogue of the exponent.
The Mittag-Leffler function 
\[ E_{\alpha}[z] =\sum^{\infty}_{k=0} \frac{z^n}{\Gamma(k \alpha+1)}  \]
is invariant with respect to the Riemann-Liouville
\[ D^{\alpha}_{a+} E_{\alpha}[\lambda (x-a)^{\alpha}] =
\lambda E_{\alpha}[\lambda (x-a)^{\alpha}] \]
(see equation 2.1.57 in \cite{KST}).


\section{Introduction to fractional dynamics}

Fractional dynamics is a field in physics and mechanics, 
studying the behavior of objects and systems 
that are described by using integrations and differentiation of fractional orders, i.e. 
by methods of fractional calculus. 
Derivatives and integrals of non-integer orders are used 
to describe objects that can be characterized by the following properties.

1) A power-law non-locality; 

2) A power-law long-term memory; 

3) A fractal-type property.


\subsection{Fractional diffusion-wave equation}

In mathematics and physics there are well-known the following equations:
the diffusion equation
\[ \nabla^2 u(t,{\bf x}) = C_1 \,  D^1_t u(t,{\bf x}) , \]
and the wave equation
\[ \nabla^2 u(t,{\bf x}) = C_2 \,  D^2_t u(t,{\bf x}) . \]
We can consider a generalization of these equations such that 
it includes derivatives of non-integer order with respect to time.
This generalization describes phenomena that can be characterized 
by diffusion and waves properties.
The fractional diffusion-wave equation 
is the linear fractional differential equation 
obtained from the classical diffusion or wave equations 
by replacing the first- or second-order time derivatives 
by a fractional derivative (in the Caputo sense) 
of order $\alpha$ with $0<\alpha<2$,
\[ \nabla^2 u(t,{\bf x}) = C_{\alpha} \,  _0^CD^{\alpha}_t u(t,{\bf x}) . \]
The solutions of these fractional partial differential equations are described in the book
\cite{KST} (see Section 6.1.2).
This equation describes diffusion-wave phenomena \cite{Main1,MLP}, 
which is also called the anomalous diffusion such that we have 
the superdiffusion for $1<\alpha<2$, and subdiffusion for $0<\alpha<1$. 
A more detailed description of these effects and phenomena 
can be found in the reviews \cite{MK,Zaslavsky1}.


\subsection{Viscoelastic material}

If the force is immediately relaxed, then the deformation disappears. 
This property is called the elasticity. 
The elasticity is a physical property of materials 
which return to their original shape after they are deformed.
The other well-known property is called the viscosity. 
The viscosity of a fluid is a measure of its 
resistance to gradual deformation by shear stress or tensile stress.

Mechanically, this behavior is represented with a spring 
of modulus $E$, which describes the instantaneous elastic response.
The stress $\sigma(t)$  is proportional to the zeroth 
derivative of strain $\varepsilon (t)$ 
for elastic solids and to the first derivative of strain for viscous fluids.

The elastic solids are described by the Hooke's law:
\[ \sigma(t)= E \ \varepsilon (t) , \]
where $E$ is the elastic moduli.

The viscous fluids are described by the law of Newtonian fluids:
\[ \sigma(t)= \eta \ D^1_t \varepsilon (t) , \]
where $\eta$ is the coefficient of viscosity.

There are materials that demonstrate these two properties 
(elasticity and viscosity) at the same time.
These materials are called viscoelastic.
To describe the fractional viscoelasticity 
G. W. Scott Blair (1947) uses the relation \cite{Scott_Blair,Gerasimov}
\[ \sigma(t)= E_{\alpha} \ _0^CD^{\alpha}_t \varepsilon (t) , \]
where $E_{\alpha}$ is a constant.

If $F(x)$ is an acting force and $x$ is the  displacement, 
then Hooke's model of elasticity 
\[ F(x) = - k D^0_tx(t) ,  \] 
and Newton's model of a viscous fluid 
\[ F(x) = - k D^1_t x(t) , \] 
can be considered as particular cases of the relation
\[ F(x) = - k D^{\alpha}_t x(t) .  \] 
This force describe the property that is called fractional friction.

More complicated fractional models for viscoelasticity of materials 
are considered in the books by Rabotnov \cite{Rabotnov} and Mainardi \cite{Mainardi}.


\subsection{Power-law memory and fractional derivatives}


A physical interpretation of equations with 
derivatives and integrals of non-integer order
with respect to time is connected with the memory effects. 

Let us consider the evolution of a dynamical system 
in which some quantity $A(t)$ is related 
to another quantity $B(t)$ through a memory function $M(t)$ by 
\be \label{4-convol} 
A(t)= \int^t_0 M(t-\tau) B(\tau) d \tau.  
\ee
This operation is a particular case of composition products 
suggested by Vito Volterra.
In mathematics, equation (\ref{4-convol}) means that
the value $A(t)$ is related with $B(t)$ by the convolution 
$A(t)=M(t)*B(t)$. 

Equation (\ref{4-convol}) is a typical equation 
obtained for the systems coupled to an environment, 
where environmental degrees of freedom being averaged. 
Let us note the memory functions for the case of 
the absence of the memory and the case of power-law memory.

 
The absence of the memory:
For a system without memory, the time dependence 
of the memory function is 
\be \label{4-Md} 
M(t-\tau)=M(t) \, \delta(t-\tau) , \ee
where $\delta(t-\tau)$ is the Dirac delta-function. 
The absence of the memory means that the function 
$A(t)$ is defined by $B(t)$ at the only instant $t$. 
In this case, the system loses all its 
values of  quantity except for one: $A(t)= M(t) B(t)$. 
Using (\ref{4-convol}) and (\ref{4-Md}), we have 
\be \label{4-delta}A (t) = 
\int^t_0 M(t) \delta(t-\tau) B(\tau) d\tau  = M(t) B(t) . 
\ee
Expression (\ref{4-delta}) corresponds to the well-known 
physical process with complete absence of memory. 
This process relates all subsequent values to previous values 
through the single current value at each time $t$. 

Power-law memory:
The power-like memory function is defined by
\be \label{4-Mpl}
M(t-\tau) =M_0 \, (t-\tau)^{\varepsilon-1} ,
\ee
where $M_0$ is a real parameter.
It indicates the presence of the fractional derivative or integral. 
The integral representation is equivalent to 
a differential equation of the fractional order. 
Substitution of (\ref{4-Mpl}) into (\ref{4-convol}) gives
the temporal fractional integral of order $\varepsilon$:
\be \label{4-FracInt}
A(t)=\lambda I^{\varepsilon}_t B(t)=
\frac{\lambda}{\Gamma(\varepsilon)} 
\int^t_0 (t-\tau)^{\varepsilon-1} B(\tau) d\tau , \quad 0< \varepsilon<1 , \ee
where $\lambda=\Gamma(\varepsilon) M_0$. 
The parameter $\lambda$ can be regarded as the strength 
of the perturbation induced by the environment of the system.
The physical interpretation of the fractional integration
is an existence of a memory effect 
with power-like memory function.
The memory determines an interval $[0,t]$ during which 
$B(\tau)$ affects $A(t)$. 

Equation (\ref{4-FracInt}) is a special case of relation
for $A(t)$ and $B(t)$, where $A(t)$ is directly proportional to $B(t)$.
In a more general case, the values $A(t)$ and $B(t)$
can be related by the equation
\be \label{GC} 
f(A(t),M(t)*D^n_tB(t))=0 , \ee
where $f$ is a smooth function.
For dynamical systems relation (\ref{GC}) defines a memory effect. 
In this case (\ref{GC})
gives the relation $f(A(t),\ _0^CD^{\alpha}_t B(t))=0$
with Caputo fractional derivative.
Relation (\ref{GC}) is a fractional differential equation.


\section{Discrete physical systems with memory}

Discrete maps (universal, Chirikov-Taylor, rotator, 
Amosov, Zaslavsky, Henon) can be obtained from the  
correspondent equations of motion with
a periodic sequence of delta-function-type pulses (kicks).

An approximation for derivatives of these equations is not used.
This fact is used to study the evolution that is described
by differential equations with periodic kicks.

Example: The universal map without memory
\be 
x_{n+1}=x_n+p_{n+1}T , \quad p_{n+1}=p_n - KT \, G[x_n] 
\ee
is obtained from the differential equation of second order with respect to time
\be 
D^2_t x(t)+ K G [x(t)] \sum^{\infty}_{k=1} \delta (t/T-k )=0 ,
\ee
where $T=2\pi / \nu$ is the period, and $K$ is an amplitude of the pulses. \\
If $G[x]=\sin(x)$, then we have  the Chirikov-Taylor map. \\ 
For $G[x]= - x$ we have the Amosov system. \\

\subsection{Universal map with Riemann-Liouville type memory}

The Cauchy-type problem for the differential equations
\be \label{Hf1}
_0D^{\alpha}_t x(t) =- K G[x(t)] \sum^{\infty}_{k=1} \delta \Bigl(\frac{t}{T}-k \Bigr) ,
\quad 1 < \alpha \leqslant 2  \ee
with the initial conditions
\be \label{RL5c}
( \, _0D^{\alpha-1}_t x) (0+)=c_1, \quad 
( \, _0D^{\alpha-2}_t x)(0+)=( \, _0I^{2-\alpha}_t x)(0+) = c_2 ,
\ee
where $_0D^{\alpha-1}_t$ is the Riemann-Liouville derivative,
is equivalent to the map equations in the form
\be \label{N1}
x_{n+1}=\frac{T^{\alpha-1}}{\Gamma(\alpha)} 
\sum^{n}_{k=1} \, p_{k+1} V_{\alpha}(n-k+1)+
\frac{c_2 T^{\alpha-2}}{\Gamma(\alpha-1)} (n+1)^{\alpha-2} ,
\ee
\be \label{N2}
p_{n+1}=p_n-KT \, G[x_n] , \quad (1<\alpha\leqslant 2),
\ee
where $p_1=c_1$, and  
$V_{\alpha}(z)=z^{\alpha-1}-(z-1)^{\alpha-1}$, $(z\geqslant 1)$. 
The proof of this statement is gived in \cite{JPA-2009,TarasovSpringer}
If $G[x]=\sin(x)$, then we have the Chirikov-Taylor map with memory. 
For $G[x]= - x$ we have the Amosov system with memory.


\subsection{Universal map with Caputo type memory}

The Cauchy-type problem for the differential equations
\be \label{CaputoVolterra1}
D^1_t x(t) =p(t) ,
\ee
\be \label{CaputoVolterra2}
_0^CD^{\alpha-1}_t p(t) = - 
K \, G[x(t)] \sum^{\infty}_{k=1} \delta \Bigl(\frac{t}{T}- k \Bigr) , 
\quad (1 <\alpha < 2) 
\ee
with the initial conditions $x(0)=x_0$, $p(0)=p_0$,
where $_0^CD^{\alpha-1}_t$ is the Caputo derivative,
is equivalent to the map equations in the form
\be \label{E9}
x_{n+1}=x_0+p_0(n+1)T - 
\frac{KT^{\alpha}}{\Gamma(\alpha)} \sum^{n}_{k=1} \, (n+1-k)^{\alpha-1} G[x_k] ,
\ee
\be \label{E10}
p_{n+1}=p_0 - 
\frac{KT^{\alpha-1}}{\Gamma(\alpha-1)} \sum^{n}_{k=1} \, (n+1-k)^{\alpha-2} G[x_k] .
\ee
The proof of this statement is gived in \cite{JPA-2009,TarasovSpringer}
This map allows us to describe fractional maps 
with memory for dynamics with usual initial conditions.


\subsection{Kicked damped rotator map with memory}

Equation of motion
\be \label{fdr} _0D^{\alpha}_t x - q \, _0D^{\beta}_t x =
K G[x] \sum^{\infty}_{n=0} \delta (t -n T ) , \ee
where $q \in \mathbb{R}$, $1< \alpha\leqslant 2$, $\beta=\alpha-1$, 
and $ _0D^{\alpha}_t$ is Riemann-Liouville derivative, 
can be represented in the form of the discrete map
\be \label{xxnappna}
x_{n+1}=\frac{1}{\Gamma(\alpha-1)} 
\sum^{n}_{k=0} p_{k+1} W_{\alpha}(q,T,n+1-k) , \ee
\be p_{n+1}=e^{qT} \Bigl( p_n+K G[x_n] \Bigr) , \ee
where the function $W_{\alpha}$ is defined by
\be \label{Wa} W_{\alpha}(q,T,m+1) = T^{\alpha-1} 
\int^{1}_0 \, e^{-qTy} \, (m+y)^{\alpha-2} \, d y . \ee
The proof of this statement is given in \cite{TarasovSpringer} (see also \cite{JMP-2009,TE}.

\subsection{New type of attractors}

The suggested maps with memory are equivalent to 
The correspondent fractional kicked differential equations 
\cite{TZ,JMP-2009,JPA-2009}.
An approximation for fractional derivatives of these equations is not used. 
This fact is used to study the evolution that is described
by fractional differential equations.
Computer simulations of the suggested discrete maps with memory
prove that the nonlinear dynamical systems,
which are described by the equations with fractional derivatives,
exhibit a new type of chaotic motion and a new type of attractors.
For example, the 
slow converging and slow diverging trajectories, ballistic trajectories, 
and fractal-like sticky attractors, in the chaotic sea 
can be observed \cite{ET} for Chirikov-Taylor map with power-law memory
(see also \cite{ET,TE,Ed-1,Ed-2,Ed-3}).


\section{Dynamics of systems with long-range interaction}


Dynamics with long-range interaction has been the subject
of continuing interest in different areas of science.
The long-range interactions have been studied
in discrete systems as well as in their continuous analogues.

The dynamics described by the equations with
fractional space derivatives can be characterized by
the solutions that have power-like tails \cite{KST}.
Similar features were observed
in the lattice models with power-like long-range
interactions \cite{Br4,GF,AEL,AK,APV}.
As it was shown \cite{JPA2006-4,JMP2006,LZ,Chaos2006-1,CNSNS2006-1,KZT,ZED,Leoncini1,Leoncini2},
the equations with fractional derivatives can be directly connected to
chain and lattice models with long-range interactions.

Equations of motion of one-dimensional lattice system of
interacting particles:
\be \label{2-Main_Eq}
\frac{d^2 u_n(t)}{d t^2} = g 
\sum^{+\infty }_{m=-\infty , m \not= n}
J(n,m) [u_n(t)-u_m(t)] + F (u_n(t)) ,
\ee
where $u_n(t)$ are displacements from the equilibrium,
$F(u_n)$ is the external on-site force, and
\be \label{2-Jnm1}
J(n,m)=J(|n-m|) , \quad
\sum^{\infty}_{n=1} |J(n)|^2 < \infty .
\ee

\subsection{Long-range interaction of power-law type}

We define a special type of interparticle interaction
\be \label{2-Aa3}
\lim_{k \to 0} 
\frac{\hat{J}_{\alpha}(k)- \hat{J}_{\alpha}(0)}{|k|^{\alpha}} 
=A_{\alpha}, \quad \alpha>0, \quad 0<|A_{\alpha}|< \infty ,
\ee
where
\be \label{2-Jak}
\hat{J}_{\alpha}(k \Delta x)=\sum^{+\infty}_{\substack{n=-\infty \\ n\not=0}} 
e^{-ikn\Delta x} J(n) = 2 \sum^{\infty}_{n=1} J(n) \cos(kn \Delta x) .
\ee
This interaction is called the interaction of power-law type $\alpha$. 

 
As an example of the power-law type interaction, we can consider 
\be \label{Jn-m} J(n-m)=\frac{1}{|n-m|^{\beta+1}} . \ee
The other examples of power-law type interaction 
are considered in \cite{JPA2006-4,JMP2006,TarasovSpringer}.

Equations of motion (\ref{2-Main_Eq}) with the power-law interaction (\ref{Jn-m})
give the following equations in the continuous limits:

\noindent
1) For $0<\beta<2$ ($\beta \not=1$) we get
the Riesz fractional derivative $D^{\alpha}_{x}$ of order $\alpha=\beta$: 
\be \label{2-E3b}
\frac{\partial^2}{\partial t^2} u(x,t) -
G_{\alpha} A_{\alpha} D^{\alpha}_{x} u(x,t) =
F\left( u(x,t) \right) , \quad 
0 < \alpha < 2, \quad (\alpha \not=1) .
\ee
2) For $\beta>2$ ($\beta \not=3,4,5,...$) we get derivative of order $\alpha=2$: 
\be \label{2-E4}
\frac{\partial^2}{\partial t^2} u(x,t) +
G_{\alpha} \zeta (\alpha -1) D^2_x u(x,t) =
F\left( u(x,t) \right) , \quad 
\alpha >2 , \quad (\alpha \not=3,4,...) ,
\ee
where $G_{\alpha}=g  |\Delta x|^{min\{\alpha;2\}}$ is the finite parameter.

\noindent
3) For $\beta=1$ we get derivative of order $\alpha=1$ and 
$\alpha=2$ for $\beta=3,5,7...$:
\be \label{2-E1}
\frac{\partial^2  u(x,t)}{\partial t^2} -
i G_{1} \; \frac{\partial u(x,t)}{\partial x} = F\left( u(x,t) \right) ,
\ee
where $G_1=\pi g  \Delta x$ is the finite parameter.

\noindent
4) For $\beta=3,5,7...$ ($\beta=2m-1$, where $m=2,3,4,...$), 
we get equation with derivative of order $\alpha=2$:
\be \label{2-E2}
\frac{\partial^2 u(x,t)}{\partial t^2} -
G_{2} \; \frac{\partial^2 u(x,t)}{\partial x^2} =
F\left( u(x,t) \right),
\ee
where
\be 
G_2=\frac{(-1)^{m-1} (2 \pi)^{2m-2}}{4(2m-2)!}  
B_{2m-2} \; g  (\Delta x)^2 \ee
are the finite parameters, and $B_{2m-2}$ are Bernoulli numbers. 

\noindent
5) For $\beta=2k-2$, where $k \in \mathbb{N}$, we have the logarithmic poles. \\

The effects of synchronization, breather-type and solution-type solutions
for the systems with nonlocal interaction of power-law type  
$0<\beta<2$ ($\beta \not=1$) were investigated \cite{Chaos2006-1,CNSNS2006-1,KZT,ZED}. 
Nonequilibrium phase transitions in the thermodynamic limit 
for long-range systems are considered in \cite{Leoncini1}.
Statistical mechanics and dynamics of solvable models 
with long-range interactions are discussed in \cite{Campa}.
Stationary states of fractional dynamics of 
systems with long-range interactions are discussed in \cite{Leoncini2}.
Fractional dynamics of systems with long-range space 
interaction and temporal memory is also considered in \cite{ZED,Leoncini2}.

Fractional derivatives with respect to coordinates 
describe power-law nonlocal properties of the distributed system. 
Therefore the fractional statistical mechanics can be considered 
as special case of the nonlocal statistical mechanics \cite{Vlasov}.
As shown in the articles \cite{JMP2006,JPA2006-4} the spatial fractional derivatives 
are connected with long-range interparticle interactions.
We prove that nonlocal alpha-interactions between particles of crystal lattice 
give continuous medium equations 
with fractional derivatives with respect to coordinates.
In the monographs by Vlasov \cite{Vlasov}, 
a nonlocal statistical model of crystal lattice is suggested. 
Therefore we conclude \cite{TarasovSpringer} that the nonlocal and fractional 
statistical mechanics are directly connected with statistical dynamics of 
systems with long-range interactions \cite{Campa}.


\subsection{Nonlocal generalization of the Korteweg-de Vries equation}

The Korteweg-de Vries equation is used in a wide range of 
physics phenomena, especially those exhibiting shock waves, 
travelling waves and solitons \cite{Miles}. 
In the quantum mechanics certain physical phenomena 
can be explained by Korteweg-de Vries models. 
This equation is used in fluid dynamics, aerodynamics, and continuum 
mechanics as a model for shock wave formation, solitons, 
turbulence, boundary layer behavior and mass transport. 

The continuous limits of the equations of lattice oscillations
\[ \frac{\partial u_n(t)}{\partial t}=g_{1}             
\sum^{+\infty }_{\substack{m=-\infty \\ m \not= n}}
J_1(n,m) [u^2_n-u^2_m] +
g_{3} \sum^{+\infty }_{\substack{m=-\infty \\ m \not= n}}
J_3(n,m) [u_n-u_m] , \]
give the nonlocal generalization of the Korteweg-de Vries equation
\[ \frac{\partial}{\partial t} u(x,t)-
G_1 u(x,t)\frac{\partial}{\partial x} u(x,t)+
G_3 \frac{\partial^3}{\partial x^3} u(x,t)=0 . \]
in the form
\be \label{KdV-1} \frac{\partial}{\partial t} u(x,t)-
G_{\alpha_1} u(x,t)\frac{\partial^{\alpha_1}}{\partial |x|^{\alpha_1}} u(x,t)+
G_{\alpha_3} \frac{\partial^{\alpha_3}}{\partial |x|^{\alpha_3}} u(x,t)=0 , \ee
where $G_{\alpha_1}=g_1 |\Delta x|^{\alpha_1}$ and
$G_{\alpha_3}=g_3 |\Delta x|^{\alpha_3}$ are finite parameters. 
Equation (\ref{KdV-1}) is a fractional generalization of Korteweg-de Vries equation \cite{Momani}.

The nonlinear power-law type interactions defined by
$f(u)=u^2$ and $f(u)=u-g u^2$ for
the discrete systems are used to derive the Burgers and Boussinesq equations
and their fractional generalizations in the continuous limit.
Note that a special case of this equation is suggested in \cite{Burger2,Wang}.  


\subsection{Nonlocal generalization of the Burgers and Boussinesq equations}

Let us consider examples of quadratic-nonlinear 
long-range interactions \cite{JPA2006-4,JMP2006,TarasovSpringer}. 

We can consider the discrete systems that are described by the equations
\be 
\frac{\partial u_n(t)}{\partial t}=g_{1}             
\sum^{+\infty }_{\substack{m=-\infty \\ m \not= n}}
J_1(n-m) [u^2_n-u^2_m] +
g_{2} \sum^{+\infty }_{\substack{m=-\infty \\ m \not= n}}
J_2(n-m) [u_n-u_m] ,
\ee
where $J_1(n-m)$ and $J_2(n-m)$ define interactions of power-law type 
with $\alpha_1$ and $\alpha_2$. 
If $\alpha_1=1$ and $\alpha_2=2$, then we get
the well-known Burgers equation that is a nonlinear partial 
differential equation, which is used to describe boundary layer behavior, 
shock wave formation, and mass transport.
If $\alpha_2=\alpha$, then we get the fractional Burgers equation 
that is suggested in \cite{Burger2}.  
In the general case, the continuous limit gives the fractional Burgers equation in the form
\be
\frac{\partial}{\partial t} u(x,t)+
G_{\alpha_1} u(x,t)\frac{\partial^{\alpha_1}}{\partial |x|^{\alpha_1}} u(x,t)-
G_{\alpha_2} \frac{\partial^{\alpha_2}}{\partial x^{\alpha_2}} u(x,t)=0 . 
\ee

We can consider the chain and lattice equations of the form
\be 
\frac{\partial^2 u_n(t)}{\partial t^2}=
g_{2} \sum^{+\infty }_{\substack{m=-\infty \\ m \not= n}}
J_2(n,m) [f(u_n)-f(u_m)] +
g_{4} \sum^{+\infty }_{\substack{m=-\infty \\ m \not= n}}
J_4(n,m) [u_n-u_m] ,
\ee
where  $f(u)=u-g  u^2$,
and $J_2(n-m)$ and $J_4(n-m)$ define the interactions of power-law types
with $\alpha_2$ and $\alpha_4$.
If $\alpha_2=2$ and $\alpha_4=4$, then in continuous limit we obtain 
the well-known nonlinear partial differential equation of forth order that is called
the Boussinesq equation. This equation was subsequently applied to problems 
in the percolation of water in porous subsurface strata, and it used to describe 
long waves in shallow water and in the analysis of many other physical processes. 
In the general case, the continuous limit gives the fractional Boussinesq equation of the form
\be 
\frac{\partial^2}{\partial t^2} u(x,t)-
G_{\alpha_2} \frac{\partial^{\alpha_2}}{\partial x^{\alpha_2}} u(x,t)+
gG_{\alpha_2} \frac{\partial^{\alpha_2}}{\partial x^{\alpha_2}} u^2(x,t)+
G_{\alpha_4} \frac{\partial^{\alpha_4}}{\partial x^{\alpha_4}} u(x,t)=0 .
\ee
Fractional generalization of Korteweg-de Vries, Burgers, Boussinesq equations 
can be used to describe  properties of media with nonlocal interaction of particles.


\section{Fractional models of fractal media}


Fractals are measurable metric sets with a non-integer 
Hausdorff dimension \cite{Fractal1,Fractal2}. 
The main property of the fractal is non-integer 
Hausdorff dimension that should be observed on all scales. 
In real physical objects the fractal structure cannot be observed on all
scales but only those for which $R_0 < R < R_m$, where
$R_0$ is the characteristic scale of the particles,
and $R_m$ is the scale of objects.
Real fractal media can be characterized by 
the asymptotic form for the relation between
the mass $M(W)$ of a region $W$ of fractal medium,
and the radius $R$ containing this mass:
\[ M_D(W) =M_0 (R/ R_0)^D , \quad R/R_0 \gg 1 . \]
The number $D$ is the mass dimension. 
The parameter $D$, does not depend on the shape of the region $W$,
or on whether the packing of sphere of radius $R_0$ is close packing,
a random packing or a porous packing with a uniform distribution of holes.

The fractality of medium means than the mass of fractal homogeneous medium
in any region $W \subset \mathbb{R}^n$
increases more slowly than the n-dimensional volume of this region:
\[ M_D(W) \sim \Bigl( V_n(W) \Bigr)^{D/n} . \]
As a result, we can define that fractal medium is a system or medium 
with non-integer physical (mass, charge, particle, ...) dimension. 

 
To describe fractal media by fractional continuous model,
we can use two different notions 
such as density of states $c_n(D,{\bf r})$
and distribution function $\rho({\bf r})$.

(1) The function $c_n(D,{\bf r})$ is a density of states 
in the $n$-dimensional Euclidean space $\mathbb{R}^n$.
The density of states describes
how closely packed permitted states of particles in the space $\mathbb{R}^n$.
The expression $c_n(D,{\bf r})\, dV_n$
represents the number of states (permitted places) between $V_n$ and $V_n +dV_n$.

(2) The function $\rho({\bf r})$ is a distribution function
for the $n$-dimensional Euclidean space $\mathbb{R}^n$.
The distribution function describes a distribution of physical values
(for example, the mass, probability, electric charge, number of particles)
on a set of possible states in the space $\mathbb{R}^n$.

Note that some elementary models of fractal density of states and 
fractal distributions by open and closed boxes are suggested in \cite{TarasovSpringer}.

\subsection{Homogeneity and fractality}

To describe the fractal medium, we use a continuous medium model.
In this model the fractality and homogeneity properties
can be realized in the following forms:

(1) Homogeneity: 
The local density of homogeneous fractal medium can be described
by the constant density $\rho({\bf r})=\rho_0=const$.
This property means that the equations with constant density
must describe the homogeneous media, i.e.,
if $\rho({\bf r})=const$ and $V(W_1)=V(W_2)$, then $M_D(W_1)=M_D(W_2)$.

(2) Fractality:
The mass of the ball region $W$ of fractal homogeneous medium
obeys a power law relation $M \sim R^{D}$,
where $0<D<3$, and $R$ is the radius of the ball.
If $V_n(W_1)= \lambda^n V_n(W_2)$ and $\rho({\bf r},t)=const$,
then the fractality means that $M_D(W_1)=\lambda^D M_D(W_2)$.

These two conditions cannot be satisfied if the mass of a medium
is described by integral of integer order.
These conditions can be realized by the fractional equation
\be \label{1-MW3}
M_D(W,t)=\int_W \rho({\bf r},t) d V_D , \quad
dV_D= c_3(D,{\bf r}) \, dV_3 ,
\ee
where ${\bf r}$ is dimensionless vector variable.


\subsection{Balance equations for fractal media}

The equation of continuity (mass balance) for fractal media
\be \label{1-1eq} 
\left(\frac{d}{dt}\right)_D \rho=-\rho \nabla^D_k u_k . \ee
The equation of momentum balance for fractal media
\be \label{1-2eq} 
\rho \left(\frac{d}{dt}\right)_D u_k=
\rho f_k+\nabla^D_l p_{kl} . \ee
The equation of energy balance for fractal media
\be \label{1-3eq} 
\rho\left(\frac{d}{dt}\right)_D e = 
c^{-1}_3(D,{\bf r}) c_2(d,{\bf r})  \, p_{kl} \, \nabla_l u_k + \nabla^D_k q_k . \ee
Here $D$ is a mass dimension of fractal medium, and
\[ \left(\frac{d}{dt}\right)_D= \frac{\partial}{\partial t}+
c^{-1}_3 (D,{\bf r}) c_2(d,{\bf r}) u_l \nabla_l , \quad
\nabla^D_k A= c^{-1}_3(D,{\bf r}) \nabla_k (c_2(d,{\bf r}) A) . \]
These equations are proved in \cite{AP2005-2}.

\subsection{Moment of inertia for fractal bodies}

The moments of inertia of fractal-homogenous rigid ball 
\be \label{1-IS2}
I^{(D)}_z=\frac{2D}{3(D+2)} M_D R^2 , \quad
\frac{I^{(D)}_z}{I^{(3)}_z} = 1+\frac{2(D-3)}{3(D+2)} .
\ee
The moments of inertia of fractal-homogenous rigid cylinder
\be \label{1-I-F} 
I^{(\alpha)}_z=\frac{\alpha}{\alpha+2} M_D R^2 , \quad
\frac{I^{(\alpha)}_z}{I^{(q)}_z} = 1+\frac{\alpha-2}{\alpha+2} .
\ee
The parameter $\alpha$ is a fractal mass dimension of the cross-section
of cylinder ($1<\alpha \leqslant 2$). 
This parameter can be easy calculated from
the experimental data by box counting method
for the cross-section of the cylinder. 


The periods of oscillation for the
Maxwell pendulum with fractal rigid cylinder
\be \label{3-TT-alpha}
\frac{T^{(\alpha)}_0}{T^{(2)}_0} = 
\sqrt{ \frac{ 4(\alpha+1) }{ 3(\alpha+2)} } . \ee
The deviation $T^{(\alpha)}_0$ from $T^{(2)}_0$ for $1<\alpha \leqslant 2$ 
is no more that 6 percent.
Equation (\ref{3-TT-alpha}) allow us to use an experimental determination
of a fractal dimensional for fractal rigid body
by measurements of periods of oscillations.


For a ball with mass $M_D$, radius $R$,
and a mass fractal dimension $D$, we can consider
the motion without slipping on an inclined plane 
with a fixed angle $\beta$ to the horizon.
The condition of rolling without slipping means that
at each time point of the ball regarding the plane
is stationary and the ball rotates on its axis.
The center of mass of a homogeneous cylinder moves in a straight line.
Using the law of energy conservation, we obtain the equation
\be
v(D)= \frac{3(D+2)}{5D+6} \, gt \, \sin \beta .
\ee  
As a result, we have 
\be
\frac{v(D)}{v(3)}= \frac{21(D+2)}{5(5D+6)} .
\ee
Note that the deviation of velocity $v(D)$ of fractal solid sphere
from the velocity $v(3)$  of usual ball is less than 5 percent.

The suggested equations allows us to measure experimentally the
fractional mass dimensions $D$ of fractal materials 
by measuring the velocities.
Note that this measured dimension $D$ must be related to the fractal dimension 
that can be determined by the box counting method.


\subsection{Dipole and quadrupole moment of charged fractal distributions}

The fractional model can be used to describe 
fractal distribution of charges \cite{MPLB2005-2,POP2005}.
The distribution of charged particles is called a homogeneous one
if all regions $W$ and $W^{\prime}$ with the equal 
volumes $V_D(W)=V_D(W^{\prime})$ have the equal total charges on 
these regions,  $Q_D(W)=Q_D(W^{\prime})$. 
For charged particles that are distributed homogeneously 
over a fractal with dimension $D$, the electric charge $Q$ satisfies the 
scaling law $Q(R) \sim R^{D}$,
whereas for a regular $n$-dimensional Euclidean object 
we have $Q(R)\sim R^n$. 
This property can be used to measure the 
fractal dimension $D$ of fractal distributions of charges.
We consider this power-law relation as a definition of
a fractal charge dimension.
If all particles of a distribution are identical, then
the charge dimension is equal to the mass dimension.
In general, these dimensions can be considered as 
different characteristics of fractal distribution. 


Let us consider the example of electric dipole moment
for the homogeneous ($\rho({\bf r})=\rho_0$) fractal distribution
of electric charges in the parallelepiped region
\be \label{1-paral} 
 0 \le x \le A,\quad  0 \le y \le B , \quad 0 \le z \le C  . \ee
In the case of Riemann-Liouville fractional integral, 
we have the dipole moment $p^{(D)}_x$ in the form
\[ p^{(D)}_x=\frac{\rho_0}{\Gamma^3(a)} 
\int^A_0 dx \int^B_0 dy \int^C_0 dz \  x^{a}y^{a-1}z^{a-1}= \]
\be
=\frac{\rho_0 (ABC)^a}{\Gamma^3(a)}  \frac{A}{a^2(a+1)} ,\ee
where $a=D/3$. 
The electric charge of parallelepiped region is defined by
\be Q(W)=\rho_0 \int_W dV_D=
\frac{\rho_0 (ABC)^{a}}{a^3 \Gamma^3(a)} . \ee
Then the dipole moment for fractal
distribution in parallelepiped is
\be p^{(D)}_x=\frac{a}{a+1} Q(W) A ,\ee
By analogy with this equation, 
\be p^{(D)}_y=\frac{a}{a+1} Q(W) B, \quad 
p^{(D)}_z=\frac{a}{a+1} Q(W) C . \ee
Using $a/(a+1)=D/(D+3)$, we obtain
\be  p^{(D)}=\frac{2 D}{D+3} p^{(3)} , \ee
where $p^{(3)}=|{\bf p}^{(3)}|$ are the dipole moment for the usual 
3-dimensional homogeneous distribution. For example, 
the relation $2\le D \le 3$ leads us to the inequality
\be 0.8 \le p^{(D)}/ p^{(3)} \le 1 . \ee
These inequalities describe dipole moment 
of fractal distribution of charged particles in the parallelepiped region.


The example of electric quadrupole moment
for the homogeneous ($\rho({\bf r})=\rho_0$) fractal distribution
in the ellipsoid region $W$:
\be\label{1-ell} 
\frac{x^2}{A^2}+\frac{y^2}{B^2}+\frac{z^2}{C^2} \le 1  \ee
is considered in \cite{TarasovSpringer,MPLB2005-2,POP2005}.
The fractional model of fractal media gives \cite{MPLB2005-2,POP2005} 
the electric quadrupole moments for fractal ellipsoid
\be Q^{(D)}_{kk}= \frac{5D}{3D+6}Q^{(3)}_{kk}, \ee
\be 
Q^{(D)}_{kl}=\frac{5 \pi}{D+2} 
\frac{\Gamma^2(D/6+1/2)}{\Gamma^2(D/6)} Q^{(3)}_{kl} ,
\ee
where $k\not=l$ amd $k,l=1,2,3$. 
For $2<D<3$, we get
\be \frac{5}{6}  < Q^{(D)}_{kk} /Q^{(3)}_{kk} < 1 , \ee
\be 0.6972 < Q^{(D)}_{kl}/Q^{(3)}_{kl} < 1 . \ee
These inequalities describe values of the diagonal and non-diagonal elements 
of the  electric quadrupole moments 
for fractal distribution of charged particles in ellipsoid region.


\subsection{Some applications of fractional models of fractal media}

In this section, we considered some fractional models 
to describe dynamics of fractal media. 
In general, the fractal medium cannot be considered as a continuous medium.
There are points and domains that are not filled of particles.
We consider the fractal media as special continuous media.
We use the procedure of replacement of the medium 
with fractal mass dimension by some continuous model 
that uses the fractional integrals.
The main notions that allows us to describe fractal media are 
a density of states and density of distributions.
The fractional integrals are used to take into account 
the fractality of the media.
Note that fractional integrals can be considered as integrals 
over the space with fractional dimension up to numerical factor \cite{TarasovSpringer}. 

The suggested fractional models of fractal media 
can have a wide application. 
This is due in part to the relatively small numbers of parameters 
that define a fractal medium of great complexity and rich structure.
In many cases, the real fractal structure of matter 
can be disregarded and we can describe the medium by a fractional model,
in which the fractional integration is used.
The order of fractional integral is equal 
to the fractal physical dimension of the medium.

The fractional continuous model allows us
to describe dynamics of fractal media and fractal distributions 
\cite{AP2005,TarasovSpringer,TarasovRCD}. 
Applications of fractional models  
to describe fractal distributions of charges 
are considered in \cite{POP2005,TarasovSpringer}.  
We note that gravitational field of fractal distribution 
of particles and fields can be described by 
fractional continuous models \cite{CMDA2006} (see also \cite{Calcagni}). 
Using fractional integrals, the fractional generalization of 
the Chapman-Kolmogorov equation and
the Fokker-Planck equation for fractal media are derived \cite{MPLB-2007}.
We note applications of fractional continuous models 
by Ostoja-Starzewski to the thermoelasticity \cite{Starzewski1}, 
and the thermomechanics \cite{Starzewski2},
the turbulence of fractal media \cite{Starzewski3},
the elastic and inelastic media with fractal geometries \cite{Starzewski5},
the fractal porous media \cite{Starzewski6}, and
the fractal solids \cite{Starzewski7}.
The hydrodynamic accretion in fractal media \cite{Roy1,Roy2,Roy3} 
is considered by Roy and Ray by using 
a fractional continuous model.


\section{Open system in environment}

The closed, isolated and Hamiltonian systems are idealizations 
that are not observable and therefore do not exist 
in the real world. 
As a rule, any system is always embedded in some environment and 
therefore it is never really closed or isolated. 
Frequently, the relevant environment is unobservable or it is unknown in principle. 
This would render the theory of open, non-Hamiltonian and 
dissipative quantum systems a fundamental generalization of 
the theory of closed Hamiltonian quantum systems. 
Now the open, dissipative and non-Hamiltonian quantum systems 
are of strong theoretical interest \cite{Op1}-\cite{Op7}.

\subsection{System and environment}

Let $Q$ and $P$ be the self-adjoint operators
of coordinate and momentum of the system respectively, 
and $q_k$ and $p_k$ describe those of the environment.

The Hamiltonian $H$ of the system  is
\be \label{Hs}
H_s=\frac{P^2}{2M}+V(Q).
\ee

As a model of environment,
we consider an infinite set of harmonic oscillators 
coupled to the system.
The environment Hamiltonian is
\be \label{He}
H_e=\sum^{N}_{n=1} \left( \frac{p^2_n}{2m_n}+
\frac{m_n\omega^2_n q^2_n}{2} \right) .
\ee
This model is called the independent-oscillator model,
since the oscillators do not interact with each other.

The interaction between the system and the environment will be considered in the form
\be \label{Hi}
H_i=-Q\sum^N_{n=1} C_n q_n+ Q^2 \sum^N_{n=1} \frac{C^2_n}{2m_n \omega^2_n} , 
\ee 
where $C_n$ are the coupling constants. 

Note that the total Hamiltonian $H=H_s+H_e+H_i$ 
for the case $V(Q)=(M\Omega^2/2)Q^2$ is 
the well-known Caldeira-Leggett Hamiltonian \cite{CL-1,CL-2}.

\subsection{Equations of motion for open systems}

Using total Hamiltonian $H=H_s+H_e+H_i$,
we can derive Heisenberg equations 
for the system and the environment. 
For the system we have
\be \frac{dQ}{dt}= \frac{1}{i \hbar} [Q,H]=M^{-1} P, \quad
\label{E1}
\frac{dP}{dt}=\frac{1}{i \hbar} [P,H] = -V^{\prime}(Q)+\sum^N_{n=1} 
\left(C_n q_n-\frac{C^2_n}{m_n \omega^2_n}Q \right) .
\ee
The Heisenberg equations for the environment are
\be \frac{dq_n}{dt}=\frac{1}{i \hbar} [q_n,H] =m^{-1}_n p_n, \quad
\label{E2}
\frac{dp_n}{dt}=\frac{1}{i \hbar} [p_n,H]=-m_n\omega^2_n q_n+ C_n Q .
\ee
Eliminating the operators $P$ and $p_n$, $n=1,...,N$, 
we can write equations (\ref{E1}) and (\ref{E2}) in the form
\be \label{E3}
M \frac{d^2 Q}{dt^2}+V^{\prime}(Q)=
\sum^N_{n=1} \left(C_nq_n-\frac{C^2_n}{m_n \omega^2_n} Q \right),
\ee
\be \label{E4}
m_n\frac{d^2 q_n}{dt^2}+ m_n\omega^2_nq_n= C_n Q .
\ee

The solution of operator equation (\ref{E4}) has the form
\be \label{E5}
q_n(t)=q_n(0)\cos(\omega_n t)+\frac{p_n(0)}{m_n \omega_n} \sin(\omega_n t)
+\frac{C_n}{m_n \omega_n} \int^t_0 Q(\tau) \sin \omega_n(t-\tau) \, d \tau ,
\ee
where $q_n(0)$ and $p_n(0)$ are the initial values of coordinate and 
momentum operators of the environment $n$-th oscillators.

Using solution of (\ref{E5}) we can derive the equation
\be \label{QGL}
M\frac{d^2 Q}{dt^2}+\int^t_0 {\cal M}(t-\tau) \frac{d Q(\tau)}{d \tau} 
d \tau+V^{\prime}(Q)= F(t),
\ee
where the function
\be \label{K}
{\cal M}(t)=\sum^N_{n=1} \frac{C^2_n}{m_n \omega^2_n} \cos(\omega_n t) 
\ee
is called the memory kernel. The one-parameter operator function 
\be \label{ft}
F(t)=\sum^N_{n=1} \left( 
C_n q_n(0)\cos(\omega_n t)+\frac{C_n p_n(0)}{m_n \omega_n} \sin(\omega_n t)
-\frac{C^2_n}{m_n \omega^2_n} Q(0) \cos(\omega_n t) \right)
\ee
can be interpreted as a stochastic force since 
the initial states of the environment are uncertain 
and it can be determined by a distribution 
of the average values of $q_n(0)$ and $p_n(0)$.


\subsection{Quantum dynamics with memory}

The memory function ${\cal M}(t)$ describes dissipation if ${\cal M}(t)$ 
is positive definite and decreases monotonically.
These conditions are achieved if $N\rightarrow \infty$
and if $C^2_n/(m_n\omega^2_n)$ and $\omega_n$ are
sufficiently smooth functions of the index $n$.

For $N\rightarrow \infty$, the sum in 
equation (\ref{K}) is replaced by the integral 
\be \label{Mt}
{\cal M}(t)=\frac{2}{\pi} \int^{\infty}_{-\infty} 
\frac{J(\omega)}{\omega} \cos(\omega t) d\omega ,
\ee
where $J(\omega)$ is spectral density.
We assume that the oscillator environment contains 
an infinite number of oscillators with a continuous spectrum.

For the spectral density
\be \label{C1}
J(\omega)=\frac{\pi \omega}{2}
\sum^N_{n=1} \frac{C^2_n}{m_n \omega^2_n} 
\delta(\omega -\omega_n) ,
\ee
equation (\ref{Mt}) gives the memory function (\ref{K}).
If we consider the Cauchy distribution
$J(\omega)= a/ (\omega^2+\lambda^2)$,
then equation (\ref{Mt}) gives the exponential memory kernel
${\cal M}(t)= (a / \lambda)\, e^{-\lambda t}$.


We can consider a power-law for the spectral density:
\be \label{frac}
J(\omega)=A\omega^{\beta}, \quad 0 < \beta <1, 
\ee
where $A>0$. Note that density (\ref{frac}) leads to the power-law 
for the memory function ${\cal M}(t) \sim t^{-\beta}$.
Equation (\ref{frac}) can be achieved by a different type of
combinations of coupling coefficients $C(\omega)$ 
and density  of states $g(\omega)$ 
\[ J(\omega)=\frac{\pi \omega}{2} g(\omega) C(\omega) .\]

Using the Fourier cosine-transform 
\be \label{Fct}
\int^{\infty}_0 x^{-\alpha} \cos(xy) dx=
\frac{\pi}{2 \Gamma(\alpha) \cos(\pi \alpha/2)} y^{\alpha-1} , 
\quad (0< \alpha <1) ,
\ee
we get the equation
\be \label{EOM}
M \frac{d^2 Q}{dt^2}+\frac{A}{\sin(\pi \beta/2)} \,  _0D^{\beta}_t Q 
+V^{\prime}(Q)=F(t) ,
\ee
where $\  _0D^{\beta}_t $ is the Caputo fractional derivative  
\be \label{R2}
_0D^{\beta}_t Q(t)=
\frac{1}{\Gamma(1-\beta)} 
\int^{t}_{0} \frac{ (D^1Q)(\tau) d\tau}{(t-\tau)^{\beta}},  
\quad (0<\beta<1) .
\ee

As a simple example of quantum system, which is described by (\ref{EOM}), 
we can consider the linear fractional oscillator that 
is an object of numerous investigations 
\cite{GM2,MG,M,ZSE,Stanislavsky1,Stanislavsky2,Stanislavsky3,TZ2,AHC,AHCb,AHEC,Tof1,Tof2}
because of different applications.

If $F(t)\not=0$, then equation (\ref{EOM}) can be considered as a fractional Langevin equation.
We also can consider a general quantum analogs of fractional Langevin equation
\cite{L1,L2,L3,L4,L5,L5b,L6,L6b,L7,L8}
that can be connected with the quantum Brownian motion
that is considered by Lindblad \cite{Lind2}.


As a result, we obtain a fractional differential equation for operators $Q(t)$
from the interaction between the system and 
the environment with power-law spectral density \cite{CEJP-2012}. 
The parameter $\alpha$ can be used to control quantum dynamics 
of nano-systems like individual atoms
and molecules in an environment \cite{Nano-4,Nano-5,Nano-6,Nano-7}. 
Quantum control is concerned with active manipulation of physical and 
chemical processes on the atomic and molecular scale. 
Controlled manipulation by atomic and molecular quantum systems 
has attracted a lot of research attention in recent years \cite{Nano-1,Nano-2,Nano-3}.
Note that the models to control of open quantum system dynamics is a 
very important subject of nanotechnology.


\subsection{Quantum analogs of fractional derivatives}

One of the way to derive a quantum description of physicsl systems
is an application of a procedure of quantization
to classical models.
We can use the Weyl quantization to obtain quantum analogs of 
differential operator of non-integer orders with respect to coordinates.

The Weyl quantization $\pi_W$ is defined by 
\be \label{22-pp1}
\pi_{W} \Bigl( q_k A(q,p) \Bigr)=
\frac{1}{2} \Bigl( Q_k \hat A +\hat A Q_k \Bigr) , \quad
\pi_{W} \Bigl( p_k A(q,p) \Bigr)=
\frac{1}{2} \Bigl( P_k \hat A +\hat A P_k \Bigr) , \ee
\be \label{22-pp3} 
\pi_{W} \Bigl( D^1_{p_k} A(q,p)\Bigr)=
-\frac{1}{i\hbar} \Bigl( P_k \hat A-\hat A P_k \Bigr), \quad
\pi_{W} \Bigl( D^1_{p_k} A(q,p)\Bigr)=
\frac{1}{i\hbar} \Bigl( Q_k \hat A-\hat A Q_k \Bigr), \ee
for any $\hat A=A(Q,P)=\pi_{W}(A(q,p))$, where $k=1,...,n$,
$Q_k=\pi_{W}(q_k)$, and $P_k=\pi_{W}(p_k)$.

Weyl quantization $\pi_W$ maps \cite{TarasovElsevier} the differential operator
${\cal L}[q,p,D^1_q,D^1_p]$ on the function space 
and the superoperator 
${\cal L}[L^{+}_Q,L^{+}_P, -L^{-}_P,L^{-}_Q]$
acting on the operator space, where
\[ L^{-}_{A} \hat B=\frac{1}{i \hbar}(\hat A \hat B- \hat B \hat A) , \quad
L^{+}_{A} \hat B=\frac{1}{2}(\hat A \hat B+ \hat B \hat A) . \]

If $A(x)$ is an analytic function for $x \in (0,b)$, then
the Riemann-Liouville fractional derivative can be represented 
in the form
\be \label{RLFD-2}
_0D^{\alpha}_x A (x)=\sum^{\infty}_{n=0} a(n,\alpha) 
x^{n-\alpha} \, D^n_x A(x) , \ee
where 
\[ a(n,\alpha) =\frac{\Gamma(\alpha+1)}{\Gamma(n+1)
\Gamma(\alpha-n+1) \Gamma(n-\alpha+1)} . \]
Equation (\ref{RLFD-2}) defines a fractional derivative on 
operator algebra \cite{TarasovElsevier}.


The Weyl quantization of the Riemann-Liouville fractional derivatives 
with respect to phase-space coordinates \cite{JMP2008-Q} gives 
\be \label{Fder5} 
_0{\cal D}^{\alpha}_{Q_k} = \pi(\, _0D^{\alpha}_{q_k})=
\sum^{\infty}_{n=0} a(n,\alpha) \,
(L^{+}_{Q_k})^{n-\alpha} \, (-L^{-}_{P_k})^n , \ee
\be \label{Fder6}
_0{\cal D}^{\alpha}_{P_k} = \pi(\, _0D^{\alpha}_{p_k})=
\sum^{\infty}_{n=0} a(n,\alpha) \,
(L^{+}_{P_k})^{n-\alpha} \, (L^{-}_{Q_k})^n , \ee
where $L^{\pm}_A$ are defined by the equations
\[ L^{-}_{A} \hat B=\frac{1}{i \hbar}(\hat A \hat B- \hat B \hat A) , \quad
L^{+}_{A} \hat B=\frac{1}{2}(\hat A \hat B+ \hat B \hat A) . \]

For example, we have
\[ _0{\cal D}^{\alpha}_{Q} Q^n=
\frac{\Gamma(n+1)}{\Gamma(n+1-\alpha)} Q^{n-\alpha} ,\quad
 _0{\cal D}^{\alpha}_{P} P^n=
\frac{\Gamma(n+1)}{\Gamma(n+1-\alpha)} P^{n-\alpha} ,\]
where $n \geqslant 1$, and $\alpha \geqslant 0$. 

It allows not only consistently formulate the evolution of 
such quantum systems, but also to consider the 
dynamics of a wide class of quantum systems, such as the 
non-local non-Hamiltonian, dissipative, and nonlinear systems. 
Quantum analogs of the non-local systems with regular and 
strange attractors can be described \cite{TarasovElsevier}.
The regular quantum attractors
can be considered as stationary states of non-Hamiltonian quantum systems.
The condition given by Davies \cite{Dav1}
defines the stationary state of non-Hamiltonian quantum system.
An example, where the stationary state is unique and approached by
all states for long times is considered by Lindblad \cite{Lind2} for
Brownian motion of quantum harmonic oscillator.
Spohn \cite{Spohn,Spohn2,Spohn3} derives sufficient condition 
for the existence of a unique stationary state for the 
non-Hamiltonian quantum system described by Lindblad equation.
The stationary solution of the Wigner function evolution equation
for non-Hamiltonian quantum system was discussed in \cite{AH,ISS}.
Quantum effects in the steady states of  the
dissipative map are considered in \cite{DC}.
Stationary pure states of quantum non-Hamiltonian systems are 
considered in \cite{Tarpre,Tarpla}.
For classical non-Hamiltonian systems, stationary states 
are presented in \cite{AP2005,MPLB,IJMPB2005}.


\section{Fractional generalization of vector calculus}

The fractional calculus has a long history from 1695.
The history of fractional vector calculus is not so long. 
It has little more than ten years and can be reduced 
to small number of papers (about twenty articles) (see Refs. in \cite{FVC}). 

A consistent fractional vector calculus 
is important for application in the following research directions.

1) Non-local statistical mechanics. It should be noted book by Vlasov \cite{Vlasov}
book is entirely devoted to nonlocal statistical mechanics.
The fractional derivatives in equations can be connected with 
a long-range power-law interparticle interactions in statistical mechanics \cite{Campa,JPA2006-4}.

2) Non-local electrodynamics \cite{SR,ABR} and \cite{TT}, 
where the spatial dispersion describes non-local properties of media.

3) Non-local hydrodynamics and waves propagation 
in media with long-range interaction \cite{Lighthill} 
(see also Sec 8.16 in \cite{TarasovSpringer}).


It is known that the theory of differential forms is 
very important in mathematics and physics \cite{Westenholz}.
The fractional differential forms can be interesting 
to formulate fractional generalizations of differential geometry,
including symplectic, Kahler, Riemann and affine-metric geometries. 
These generalizations allow us to derive new rigorous results 
in modern theoretical physics and astrophysics \cite{Calcagni}, 
and in fractional generalization of relativistic field theory
\cite{F01,F02,F03,F04,F04b,F05,F06,F07,F08,F09,F10,F11,F12,F13} 
in curved space-time.
We assume that the fractional differential forms and 
fractional integral theorems for these forms can also be used 
to describe classical dynamics \cite{JPA-2005} and thermodynamics. 

It is important to have fractional generalizations of 
the symplectic geometry, Lie and Poisson algebras,
the concept of derivation on operator algebras. 
It allows to apply a generalization of algebraic structures
of fractional calculus to classical and quantum mechanics.
Note that the theory of operator algebras is very important
in quantum theory \cite{Emch,TarasovElsevier}.

\section*{Acknowledgments}

I would like to express my gratitude for kind invitation 
of Professor S.C. Lim to write a review on fractional dynamics.
I acknowledge with thanks stimulating discussions with 
Prof. J.J. Trujillo and with the participants of seminars 
at the Mathematical Analysis Department of University of La Laguna (Spain)
and at the Skobeltsyn Institute of Nuclear Physics.


\end{document}